\documentclass[%
 reprint,
showpacs,preprintnumbers,
 amsmath,amssymb,
 aps,prl
]{revtex4-1}

\usepackage{graphicx}% Include figure files
\usepackage{epsfig}
\usepackage{dcolumn}% Align table columns on decimal point
\usepackage{bm}% bold math
\usepackage{newfloat}
\DeclareFloatingEnvironment[name={S}]{suppfig}
\begin{document}

%\preprint{APS/123-QED}

\title{Thermomechanical two-mode squeezing in an ultrahigh $Q$ membrane resonator}
\author{Y. S. Patil, S. Chakram, L. Chang and M. Vengalattore}
 \affiliation{Laboratory of Atomic and Solid State Physics, Cornell University, Ithaca, NY 14853}%Lines break automatically or can be forced with \\
\email{mukundv@cornell.edu}
%\affiliation{%

\date{\today}% It is always \today, today,

\begin{abstract}
We realize a nondegenerate parametric amplifier in an ultrahigh $Q$ mechanical membrane resonator and demonstrate two-mode thermomechanical noise squeezing. Our measurements are accurately described by a two-mode model that attributes this nonlinear mechanical interaction to a substrate-mediated process which is dramatically enhanced by the quality factors of the individual modes. This realization of strong multimode nonlinearities in a mechanical platform compatible with quantum-limited optical detection and cooling makes this a powerful system for nonlinear approaches to quantum metrology, transduction between optical and phononic fields and the quantum manipulation of phononic degrees of freedom. 
%We realize a parametric nonlinearity in an ultrahigh quality factor mechanical resonator and demonstrate steady-state two-mode thermomechanical squeezing. Our observations are well described by a model that ascribes this nonlinearity to a substrate-mediated interaction between diatinct mechanical modes. combination of strong nonlinearity, lrage fQ product, compatibility with optomechanical cooling to the quantm regime and quantum limited optical detection - powerful system for quantum enhanced metrology, signal transduction as well as foundational studies of the quantum to classical transition, macroscopic quantum mechanics and measurement.  
\end{abstract}
\maketitle

% MAIN MESSAGE OF INTRODUCTION : IDENTIFY AND STUDY A STRONG TWO-MODE NONLINEARITY IN AN ULTRAHIGH Q MEMBRANE RESONATOR WITH LARGE FQ PRODUCT, AMENABLE TO QUANTUM LIMITED COOLING, DETECTION AND MANIPULATION : DEMONSTRATE STRONG PARAMETRIC GAIN/DEAMPLIFICATION DUE TO THIS NONLINEARITY AND STEADY STATE SQUEEZING. 

The control, measurement and manipulation of mesoscopic mechanical resonators by coherent optical fields has garnered widespread attention in recent years for potential applications to quantum metrology as well as to foundational studies of the quantum-to-classical transition and the quantum mechanics of macroscopic objects \cite{kippenberg2008,aspelmeyer2012, meystre2013, aspelmeyer2013}. Notable accomplishments in recent years include the optical cooling of mechanical modes to the quantum regime \cite{teufel2011, chan2011} and the detection of mechanical motion with an imprecision below the standard quantum limit \cite{teufel2009, anetsberger2010}.  %the demonstration of entanglement between mechanical and optical fields \cite{} and the creation of nonclassical states of a mechanical resonator via back-action evading measurements \cite{suh2014}. 

Building upon these developments, attention has now been directed towards the creation of nonclassical mechanical states and the manipulation of phononic fields in a manner akin to quantum optics in nonlinear media. In contrast to the cooling and detection of mechanical motion, the creation of nonclassical states requires strong nonlinear interactions involving the mechanical degree of freedom. Accordingly, several studies have been devoted to the realization of such interactions through parametric processes \cite{rugar1991, natarajan1995, karabalin2011, faust2013, mahboob2014}, optically mediated nonlinearities \cite{seok2012}, reservoir engineering \cite{tan2013, woolley2014}, dispersive coupling to an auxiliary quantum system \cite{oconnell2010,suh2010}, backaction-evading measurements \cite{hertz2010,suh2014} and active feedback \cite{szork2013, vinante2013, szork2014, pontin2014}.  Notwithstanding the diversity of such schemes, it has remained a significant challenge to juxtapose strong, tunable and quantum-compatible nonlinear interactions with the stringent constraints for ground state cooling and quantum control of a mechanical resonator. 

Here, we demonstrate a strong two-mode nonlinearity in an ultrahigh quality factor membrane resonator and use this nonlinear interaction to realize nondegenerate parametric amplification and two-mode thermomechanical noise squeezing. We develop a model that ascribes this two-mode nonlinearity to a substrate-mediated interaction between distinct, optically addressable modes of the resonator and find close agreement between our observations and the predictions of this model. This combination of strong multimode mechanical nonlinearities, optical addressability of individual modes, large $f \times Q$ products \cite{chakram2014}, compatibility with optomechanical cooling to the quantum regime \cite{purdy2014, lee2014} and quantum-limited optical measurement makes this a powerful system for quantum-enhanced metrology and the quantum manipulation of phononic fields.  
%This ability to generate strong multimode nonlinearities in a mechanical platform that features large $f \times Q$ products \cite{chakram2014}, is compatible with optomechanical cooling to the quantum regime \cite{purdy2014, lee2014} and quantum-limited optical measurement makes this a powerful system for quantum-enhanced metrology and the quantum manipulation of phononic fields.  

%DESCRIPTION OF RESONATOR, GEOMETRY, QUALITY FACTORS, INFLUENCE OF SUBSTRATE (CLAMPING) ON MODES, HYBRIDIZATION AND DISSIPATION. EXTENSION TO TWO-MODE COUPLING MEDIATED BY THE SUBSTRATE - ESPECIALLY SIGNIFICANT FOR PARAMETRIC PROCESSES THAT INVOLVE A DISCRETE EXCITATION OF THE SUBSTRATE - A FORM OF RESERVOIR ENGINEERING - LEADING TO MAIN FINDINGS OF PAPER

The mechanical resonators in our study are LPCVD silicon nitride (SiN) membrane resonators manufactured by NORCADA Inc. The membranes are deposited on single crystal silicon wafers and have typical lateral dimensions of 5 mm. 
%The mechanical eigenfrequencies of these membranes are well described by thin plate theory and are given by $\omega_{jk} = 2 \pi \sqrt{\sigma/4 \rho L^2} \sqrt{j^2+k^2}$ where $\sigma$ is the intrinsic tensile strain, $\rho = 2.7$ g/cm$^3$ is the mass density and $L$ is the lateral dimension of the membrane. 
In previous work \cite{chakram2014}, we have identified the role of the substrate in inducing the hybridization of proximal eigenmodes and in modifying the dissipation of the resultant hybridized modes. This leads to the robust formation of a large number of mechanical modes with quality factors $Q \sim 50 \times 10^6$ and $f \times Q \sim 1 \times 10^{14}$ Hz. %Furthermore, these hybridized modes can be imaged and optically addressed by a dark ground imaging technique \cite{imaging2013}. 

In addition to modifying the modal geometry and dissipative properties, the substrate can also mediate and enhance nonlinear interactions between distinct eigenmodes of the resonator - a form of reservoir engineering. This is especially significant for parametric processes that involve the interaction of two membrane modes mediated by a discrete excitation of the substrate, since the coupling strength is now enhanced by the quality factor of the relevant substrate mode. In our work, this nonlinear interaction between mechanical modes at frequencies $\omega_i, \omega_j$ is induced by parametrically actuating the substrate at frequencies near $\omega_i + \omega_j$, either by a piezo-electric voltage or a photothermal modulation. 

%In this work, we describe such a two-mode interaction and demonstrate parametric amplification and thermomechanical two-mode squeezing that arise from this nonlinear coupling. 

\begin{figure*}[t]
\centering
\includegraphics[width=01.0\textwidth]{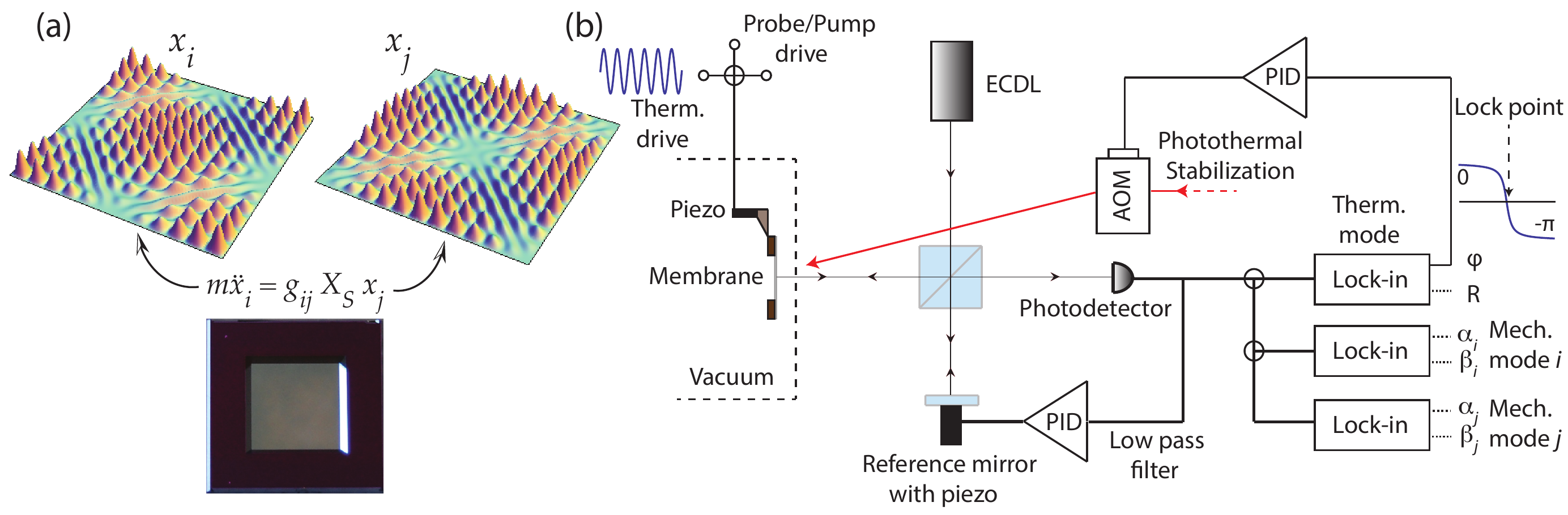}
\caption{(a) The resonator consists of a high-stress silicon nitride membrane deposited on a silicon substrate. Proximal eigenmodes of the membrane resonator (representative eigenfunctions shown) with frequencies $\omega_{i,j}$ are coupled via a substrate-mediated interaction. This two-mode interaction can be controlled by actuating the substrate to an amplitude $X_S$ at frequencies close to $\omega_i + \omega_j$. (b) Experimental schematic: Mechanical motion of the membrane is optically detected in a Michelson interferometer, with the two membrane modes $(i,j)$ distinguished by phase-sensitive lock-in detection. The eigenfrequency of a third, high-$Q$ mechanical mode is continuously monitored and acts as a mechanical `thermometer'. The resonator modes are actively frequency stabilized to this thermometer mode by photothermal feedback. In the presence of this feedback, the frequency stability of each resonator mode is better than 1 ppb over 1000 seconds.}
\label{fig:qze}
\end{figure*}
% EXPERIMENTAL SCHEMATIC : INTERFEROMETRIC DETECTION OF MECHANICAL MOTION, INTERFEROMETRIC IMAGING OF MEMBRANE MODES, OPTICAL ADDRESSABILITY, HYBRIDIZATION, PHASE SENSITIVIE DETECTION OF DIFFERENT MODES, NARROW LINEWIDTH, RESOLUTION OF THERMOMECHANICAL MOTION AND NON-CLASSICAL CORRELATIONS REQUIRES ACTIVE STABILIZATION, (SUPPLEMENTARY), BRIEF DESCRIPTION AND RESULTS THAT ENABLE THE STUDIES, PIEZO ACTUATION BUT ALSO ABLE TO SEE SIMILAR EFFECTS THROUGH PHOTOTHERMAL ACTUATION. 
A schematic of the experimental system is shown in Fig. 1. The mechanical motion of the membrane is optically detected in a Michelson interferometer with a displacement sensitivity of 0.03 pm/Hz$^{1/2}$ for typical powers of 200 $\mu$W incident on the membrane. Distinct eigenmodes are resolved through phase sensitive lock-in detection. The membrane modes exhibiting the two-mode nonlinearities studied in this work are characterized by eigenfrequencies $\omega_i/2 \pi \approx 1.5$ MHz, quality factors $Q > 10 \times 10^6$, typical mechanical linewidths $\gamma_i/2 \pi < 100$ mHz and $|\omega_i - \omega_j| > 10^6 \, \gamma_{i,j}$. While the large quality factors are crucial to realizing long coherence times, the narrow linewidths also pose stringent requirements of thermal stability in order to resolve thermomechanical motion and the presence of non-thermal correlations. To achieve the requisite frequency stability, the membrane modes are actively stabilized by photothermal feedback %to within 1 mHz over typical measurement durations of 1000 \!s 
(see Supplementary Information, see also \cite{gavartin2013}). 

The substrate-mediated coupling between a pair of membrane modes can be modeled by an interaction $\mathcal{H}_{ij} = - g_{ij} X_S x_i x_j$ where $g_{ij}$ parametrizes the strength of the two-mode coupling, $X_S$ is the displacement of the substrate (or `pump') mode and $x_{i, j}$ denotes the displacement of the two membrane modes. This interaction can be attributed to a parametric excitation of a discrete mode of the substrate at a frequency $\omega_S = \omega_i + \omega_j$, that effectively couples the membrane modes. Actuation of the substrate at this frequency thus leads to nondegenerate parametric amplification of the individual membrane modes. 

As is well known in such parametric amplifiers, a sufficiently strong interaction (or a large actuation of the pump field) leads to an instability and self-oscillation of the individual membrane modes. In the case of resonant actuation, our two-mode model predicts a threshold amplitude for self-oscillation given by (see Supplementary Information), 
\begin{equation}
X_{S, th}(g) = 2 \sqrt{\frac{1}{g^2}\times\frac{1}{\chi_i \chi_j}} \propto \sqrt{\frac{1}{g^2}\frac{1}{Q_i Q_j}}
\end{equation}
where $\chi = (m \omega \gamma)^{-1}$ are the on-resonant mechanical susceptibilities of the two membrane modes and $g$ is the strength of the two-mode coupling. The inverse dependence of the threshold pump amplitude on the quality factors of the high-$Q$ membrane modes leads to strong nonlinear behavior even for pump displacements on the order of 10 fm. Past the instability threshold, we observe amplification of the membrane modes by more than 30 dB (see Fig. 2(a)). 

\begin{figure}[t]
\includegraphics[width=2.8in]{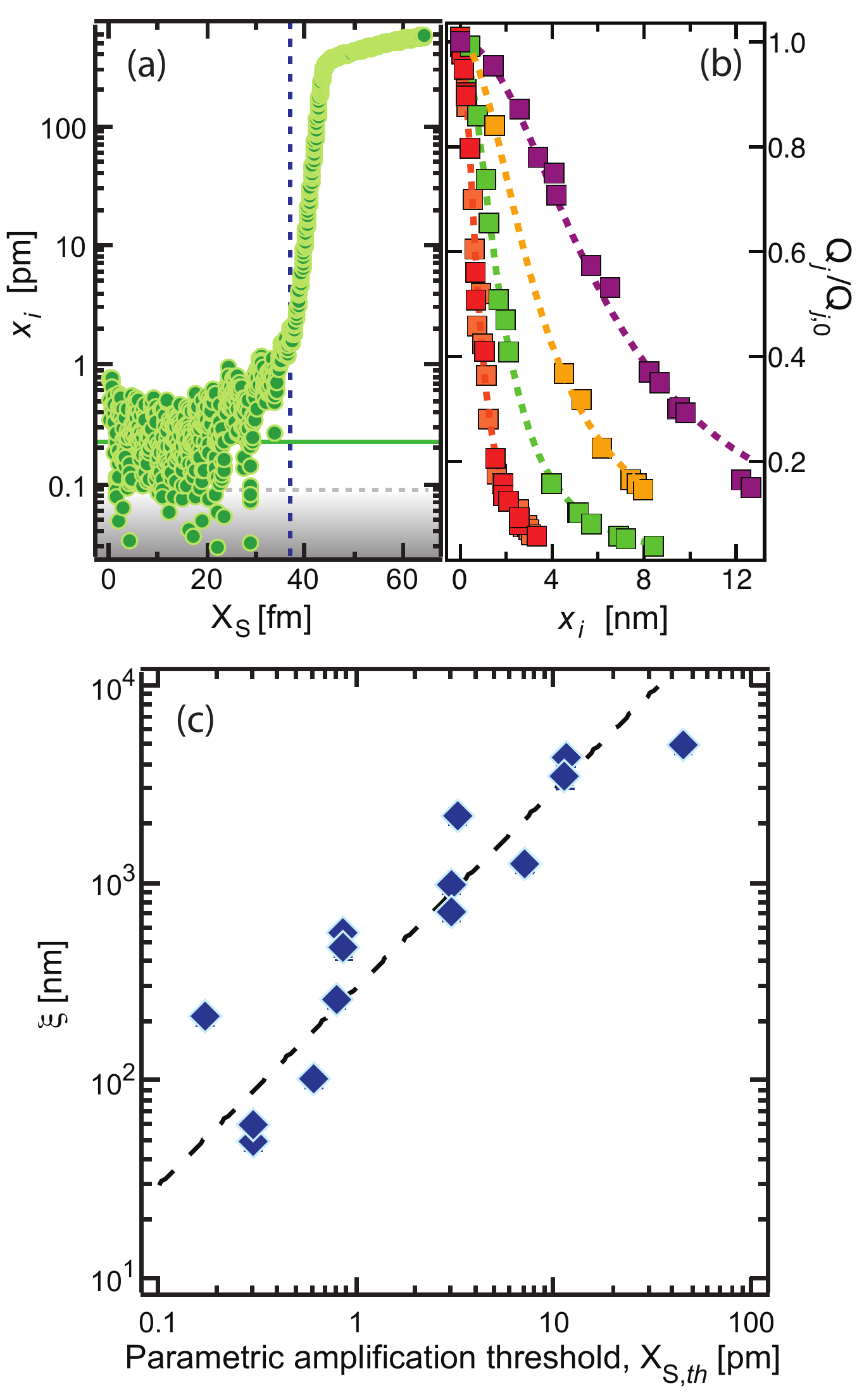}
\caption{(a) Parametric amplification of a membrane mode due to actuation of the substrate. The vertical line indicates the threshold for parametric instability. The solid green line indicates the thermomechanical amplitude of the membrane mode. The dashed grey line shows the detection noise floor. (b) In the absence of the parametric drive, large amplitude oscillations of either membrane mode ($x_i$) results in increased dissipation (and a lower quality factor $Q_j$) of the other mode due to up-conversion of excitations into the substrate. The variation of the normalized dissipation ($Q_j/Q_{j,0}$) is well described by a characteristic length scale $\xi$ (see text). (c) The linear dependence of the length scale $\xi$ {\em vs} the threshold amplitude for parametric instability, as predicted by the two-mode model. }
\label{Fig:Fig2}
\end{figure}

While the parametric amplification of the membrane modes can be regarded as a down-conversion of substrate excitations, a related process is the up-conversion of excitations from the membrane into the substrate. In the absence of substrate motion, actuation of the membrane modes can lead to coherent transfer of energy from the membrane into the substrate. From the perspective of either membrane mode, this parametric transfer of energy into the far lossier substrate results in a dissipation rate that is dependent on the amplitude of the other membrane mode. 

To study this process, the mechanical ring-down time $\tau$ of a membrane mode $j$ was measured in the presence of a large amplitude actuation of its partner mode $i$. Care was taken in these measurements to maintain the amplitude of mode $j$ in the linear response regime. The two-mode upconversion process results in an effective quality factor $Q_j(x_i) = \omega_j \tau(x_i)/2$ and an effective mechanical linewidth $\gamma_j (x_i) = 2/\tau(x_i)$ that is in excellent agreement with the prediction of the two-mode model (see Supplementary Information), 
\begin{equation}
\gamma_j(x_i)  \approx \frac{\gamma_S}{2} \left[ 1 + \frac{2 \gamma_j}{\gamma_S} - \sqrt{1 - \left( \frac{x_i}{\xi} \right)^2}\,\,\right]
\end{equation}
where $\gamma_j \ll \gamma_S$ are the intrinsic mechanical linewidths of membrane mode $j$ and the substrate mode (see Fig. 2(b)). The length scale $\xi$ denotes the characteristic amplitude of mode $i$ when the dissipation rate of the membrane mode $j$ matches that of the substrate, i.e. the maximal rate of up-conversion of energy from the membrane into the substrate. Due to the large mismatch between the intrinsic dissipation rates and masses of the substrate and the membrane, this up-conversion process requires displacements of the membrane modes that are more than five orders of magnitude larger than typical thermomechanical motion, and much larger than the typical amplitudes of motion considered in this work. %As can be seen in Fig.2(b), measurements of the nonlinear dissipation for a range of membrane mode pairs are in excellent agreement with the model. 

While seemingly distinct processes observed at vastly differing scales of displacement (fm {\em vs} nm), nondegenerate parametric amplification and nonlinear  dissipation due to up-conversion of excitations into the substrate both owe their origins to the two-mode nonlinearity. Indeed, the model predicts that the length scale $\xi$ that parametrizes two-mode control of mechanical dissipation, can be related to the threshold amplitude $X_{S,th}$ according to the relation $\xi(g) = \frac{1}{2} (\gamma_S/\gamma_i)^{1/2} (\chi_j/\chi_S)^{1/2} \times X_{S,th}(g)$. Through independent measurements of parametric amplification thresholds and nonlinear two-mode dissipation for a wide range of membrane mode pairs, we have verified this linear dependence (see Fig. 2(c)).  As can be seen, the various mode pairs that were studied exhibit interaction strengths that vary over three orders of magnitude. The close agreement between our measurements and the predictions of the model over a large dynamic range of parameters further affirms the robustness and fidelity of this nonlinear interaction in our system. 

\begin{figure}[t]
\includegraphics[width=3.00in]{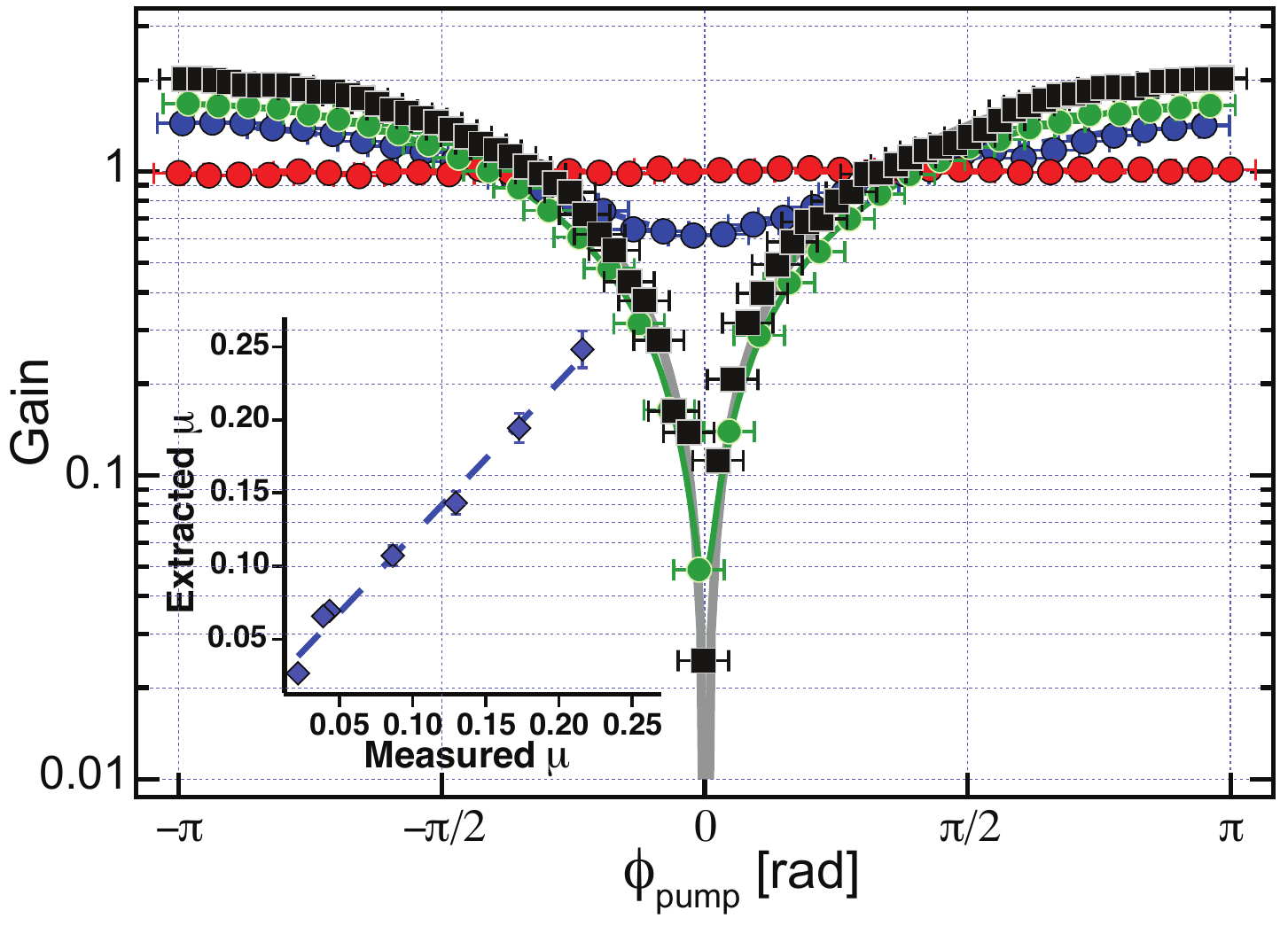}
\caption{Parametric gain versus the phase of the pump excitation. Data shown correspond to normalized pump amplitudes $\mu = X_S/X_{S,th} = 0, 0.021, 0.038, 0.042$ (red, blue, green, black). For these data, the threshold for parametric self-oscillation corresponds to $X_S = 40$ fm.  Inset: Estimate of the pump amplitude from fits to these data agree with the actual pump amplitude to within 5\%. }
\label{Fig:Fig3}
\end{figure}

Having established the accuracy of our two-mode model, we now discuss the dynamics of this nondegenerate parametric amplifier for a pump field driven below the threshold $X_{S, th}$. In this regime, weak actuation of a membrane mode $i$ (idler mode) results in the phase coherent production of down-converted phonons in the other membrane mode $j$ (signal mode). 
This down-converted field, which has a well determined phase relationship with the pump and idler fields, can coherently interfere with any pre-existing signal field. Thus, the gain of the signal field acquires a strong dependence on the pump phase $\phi$ according to the relation (see Supplementary Information)
\begin{equation}
G_j(\phi) = \frac{1}{1 - \mu^2} \sqrt{1 + \mu^2 \eta^2  - 2 \mu \eta \cos \phi} 
\end{equation}
where $\mu = X_S/X_{S,th}$ is the pump amplitude normalized to the threshold amplitude for parametric instability, and $\eta = (\chi_j/\chi_i)^{1/2} \times (\bar{x}_i/\bar{x}_j)$ where $\bar{x}_{i,j}$ are the amplitudes of the membrane modes in the absence of the pump. 

We demonstrate phase-dependent amplification by simultaneously monitoring the amplitudes of the signal and idler fields in the presence of the pump field. For these measurements, the signal and idler modes were weakly actuated while keeping their phases fixed. The pump (substrate) was actuated to different amplitudes below threshold while its phase, relative to the signal and idler, was slowly changed. The phase dependent gain of the signal mode for different values of the normalized pump amplitude $\mu$ and pump phase $\phi$ is shown in Fig. 3. For these data, the signal mode was actuated to an amplitude of $35 \times (k_B T/m \omega_j^2)^{1/2}$ while the idler mode was actuated to an amplitude of $400 \times (k_B T/m \omega_i^2)^{1/2}$. The data show excellent agreement with the above expression, with observed parametric deamplification exceeding 20 dB. The pump amplitudes extracted from fits to these data are in agreement with the independently measured amplitudes to within 5\% (Fig. 3 (inset)). 

We also note here the important distinction between the phase-dependent gain as seen in our system and that observed in a degenerate parametric amplifier. In the latter case, the maximal deamplification is limited to 0.5 (3 dB) before the onset of parametric instability. In contrast, the additional degree of freedom (the idler field) made available in the nondegenerate amplifier allows for an arbitrarily large degree of deamplification. 

% AVERAGED ACROSS DIFFERENT PHASES STILL LEADS TO SIGNIFICANT STEADY STATE GENERATION OF NON-CLASSICAL CORRELATIONS, THERMOMECHANICAL SQUEEZING. QUANTITATIVELY MODELED AND AGREES WITH CALCULATIONS. 

\begin{figure}[t]
\includegraphics[width=3.5in]{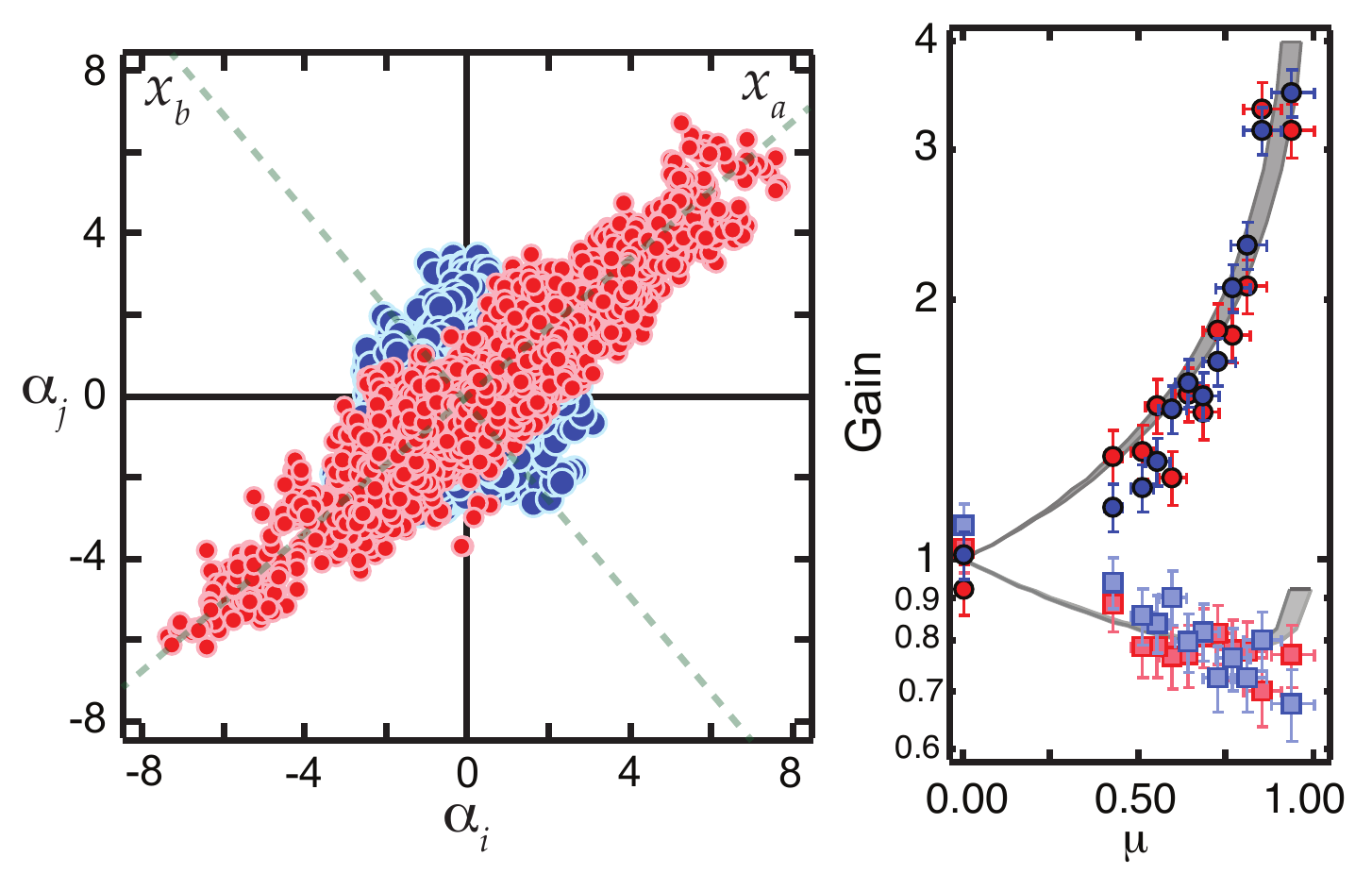}
\caption{Steady state thermomechanical two-mode squeezing. Left: Phase space distributions of the quadratures $\alpha_i, \alpha_j$ in the absence (blue) and presence (red) of the pump field, showing the emergence of correlations, i.e. noise squeezing, due to nondegenerate parametric amplification. Right: The standard deviations of the cross-quadratures $x_a, y_b$ (amplified) and $x_b, y_a$ (squeezed) {\em vs} pump amplitude. The shaded curves indicate the no-free-parameter prediction of our noise squeezing model based on independently measured parameters of our system (see SI). }
\label{Fig:Fig4}
\end{figure}

We make use of this nondegenerate parametric amplifier to demonstrate thermomechanical two-mode noise squeezing. In the absence of any actuation, the signal and idler membrane modes are subject only to thermomechanical noise. In this situation, if the pump field is driven below threshold, the membrane modes become highly correlated, with the correlations being manifest as a squeezing of a composite quadrature formed from linear combinations of quadratures of the individual membrane modes. This is the thermomechanical analog of two-mode squeezing seen in optical parametric amplifiers \cite{reid1988}. 

To quantify the degree of two-mode squeezing, we construct cross-quadratures from the displacements of the membrane modes, according to the relations $x_{a,b} = (\alpha_i \pm \alpha_j)/\sqrt{2}, y_{a,b} = (\beta_i \pm \beta_j)/\sqrt{2}$ where $\{\alpha_i, \beta_i\}$ are the respective quadratures of the individual membrane modes normalized to thermomechanical amplitudes. Phase-space distributions of these quadratures, accumulated over typical durations of 300 s, are shown in Fig. 4. The phase space distributions, which are symmetric in the absence of down-conversion, acquire a large ellipticity for increasing amplitudes of the pump field. The data are in very good agreement with our two-mode thermomechanical noise squeezing model (see Supplementary Information). While an arbitrarily large degree of squeezing may be obtained for specific phase relationships between the signal, idler and pump fields, our noise squeezing model predicts that the maximal degree of steady-state thermomechanical squeezing is limited by thermal averaging across all possible phases between the fields. Nonetheless, by harnessing weak measurements and feedback, we estimate that mechanical squeezing of more than 40 dB may be obtained with our demonstrated parameters.

% SIGNIFICANCE OF OUR WORK, BROADER PICTURE : STRONG TWO-MODE NONLINEARITIES IN A RESONATOR SYSTEM THAT IS DEMONSTRABLY COMPATIBLE WITH OPTOMECHANICAL DETECTION, COOLING AND CONTROL AT THE QUANTUM REGIME. POWERFUL EXTENSION TO THE CAVITY OPTOMECHANICS TOOLBOX OF SUCH MEMBRANE IN THE MIDDLE SYSTEMS. DEVELOPED A MODEL THAT DESCRIBES OUR FINDINGS WITH HIGH ACCURACY, LARGE DEGREE OF PHASE-SENSITIVE GAIN/DE-AMPLIFICATION - HARNESSED FOR QUANTUM-ENHANCED METROLOGY AS WELL AS FOUNDATIONAL TESTS OF MACROSCOPIC QM. 

In summary, our demonstration of nondegenerate parametric amplification in a mechanical platform featuring large $f \times Q$ products ($\gg k_B T/h$), low dissipation ($\gamma < (10 $ s$)^{-1}$) and demonstrated compatibility with cavity optomechanical cooling and quantum-limited detection, provides a powerful tool for nonlinear approaches to quantum sensing and QND measurements of mechanical degrees of freedom. In addition, our work also paves the way towards the quantum manipulation of phononic fields for studies of macroscopic entanglement.  

This work was supported by the DARPA QuASAR program through a grant from the ARO, an NSF INSPIRE award and the Cornell Center for Materials Research with funding from the NSF MRSEC program (DMR-1120296). We acknowledge experimental assistance from A. Shaffer-Moag, H. Cheung and J. Geng. M. V. acknowledges support from the Alfred P. Sloan Foundation. 

\bibliography{SiNbib, SiNnote}

\begin{thebibliography}{31}%
\makeatletter
\providecommand \@ifxundefined [1]{%
 \@ifx{#1\undefined}
}%
\providecommand \@ifnum [1]{%
 \ifnum #1\expandafter \@firstoftwo
 \else \expandafter \@secondoftwo
 \fi
}%
\providecommand \@ifx [1]{%
 \ifx #1\expandafter \@firstoftwo
 \else \expandafter \@secondoftwo
 \fi
}%
\providecommand \natexlab [1]{#1}%
\providecommand \enquote  [1]{``#1''}%
\providecommand \bibnamefont  [1]{#1}%
\providecommand \bibfnamefont [1]{#1}%
\providecommand \citenamefont [1]{#1}%
\providecommand \href@noop [0]{\@secondoftwo}%
\providecommand \href [0]{\begingroup \@sanitize@url \@href}%
\providecommand \@href[1]{\@@startlink{#1}\@@href}%
\providecommand \@@href[1]{\endgroup#1\@@endlink}%
\providecommand \@sanitize@url [0]{\catcode `\\12\catcode `\$12\catcode
  `\&12\catcode `\#12\catcode `\^12\catcode `\_12\catcode `\%12\relax}%
\providecommand \@@startlink[1]{}%
\providecommand \@@endlink[0]{}%
\providecommand \url  [0]{\begingroup\@sanitize@url \@url }%
\providecommand \@url [1]{\endgroup\@href {#1}{\urlprefix }}%
\providecommand \urlprefix  [0]{URL }%
\providecommand \Eprint [0]{\href }%
\providecommand \doibase [0]{http://dx.doi.org/}%
\providecommand \selectlanguage [0]{\@gobble}%
\providecommand \bibinfo  [0]{\@secondoftwo}%
\providecommand \bibfield  [0]{\@secondoftwo}%
\providecommand \translation [1]{[#1]}%
\providecommand \BibitemOpen [0]{}%
\providecommand \bibitemStop [0]{}%
\providecommand \bibitemNoStop [0]{.\EOS\space}%
\providecommand \EOS [0]{\spacefactor3000\relax}%
\providecommand \BibitemShut  [1]{\csname bibitem#1\endcsname}%
\let\auto@bib@innerbib\@empty
%</preamble>
\bibitem [{\citenamefont {Kippenberg}\ and\ \citenamefont
  {Vahala}(2008)}]{kippenberg2008}%
  \BibitemOpen
  \bibfield  {author} {\bibinfo {author} {\bibfnamefont {T.~J.}\ \bibnamefont
  {Kippenberg}}\ and\ \bibinfo {author} {\bibfnamefont {K.~J.}\ \bibnamefont
  {Vahala}},\ }\href@noop {} {\bibfield  {journal} {\bibinfo  {journal}
  {Science}\ }\textbf {\bibinfo {volume} {321}},\ \bibinfo {pages} {1172}
  (\bibinfo {year} {2008})}\BibitemShut {NoStop}%
\bibitem [{\citenamefont {Aspelmeyer}\ \emph {et~al.}(2012)\citenamefont
  {Aspelmeyer}, \citenamefont {Meystre},\ and\ \citenamefont
  {Schwab}}]{aspelmeyer2012}%
  \BibitemOpen
  \bibfield  {author} {\bibinfo {author} {\bibfnamefont {M.}~\bibnamefont
  {Aspelmeyer}}, \bibinfo {author} {\bibfnamefont {P.}~\bibnamefont {Meystre}},
  \ and\ \bibinfo {author} {\bibfnamefont {K.}~\bibnamefont {Schwab}},\
  }\href@noop {} {\bibfield  {journal} {\bibinfo  {journal} {Phys. Today}\
  }\textbf {\bibinfo {volume} {65}},\ \bibinfo {pages} {29} (\bibinfo {year}
  {2012})}\BibitemShut {NoStop}%
\bibitem [{\citenamefont {Meystre}(2013)}]{meystre2013}%
  \BibitemOpen
  \bibfield  {author} {\bibinfo {author} {\bibfnamefont {P.}~\bibnamefont
  {Meystre}},\ }\href@noop {} {\bibfield  {journal} {\bibinfo  {journal} {Ann.
  der Phys.}\ }\textbf {\bibinfo {volume} {525}},\ \bibinfo {pages} {215}
  (\bibinfo {year} {2013})}\BibitemShut {NoStop}%
\bibitem [{\citenamefont {Aspelmeyer}\ \emph {et~al.}(2013)\citenamefont
  {Aspelmeyer}, \citenamefont {Kippenberg},\ and\ \citenamefont
  {Marquadt}}]{aspelmeyer2013}%
  \BibitemOpen
  \bibfield  {author} {\bibinfo {author} {\bibfnamefont {M.}~\bibnamefont
  {Aspelmeyer}}, \bibinfo {author} {\bibfnamefont {T.~J.}\ \bibnamefont
  {Kippenberg}}, \ and\ \bibinfo {author} {\bibfnamefont {F.}~\bibnamefont
  {Marquadt}},\ }\href@noop {} {\bibfield  {journal} {\bibinfo  {journal}
  {arXiv:1303.0733}\ } (\bibinfo {year} {2013})}\BibitemShut {NoStop}%
\bibitem [{\citenamefont {Teufel}\ \emph {et~al.}(2011)\citenamefont {Teufel},
  \citenamefont {Donner}, \citenamefont {Li}, \citenamefont {Harlow},
  \citenamefont {Allman}, \citenamefont {Cicak}, \citenamefont {Sirois},
  \citenamefont {Whittaker}, \citenamefont {Lehnert},\ and\ \citenamefont
  {Simmonds}}]{teufel2011}%
  \BibitemOpen
  \bibfield  {author} {\bibinfo {author} {\bibfnamefont {J.~D.}\ \bibnamefont
  {Teufel}}, \bibinfo {author} {\bibfnamefont {T.}~\bibnamefont {Donner}},
  \bibinfo {author} {\bibfnamefont {D.}~\bibnamefont {Li}}, \bibinfo {author}
  {\bibfnamefont {J.~W.}\ \bibnamefont {Harlow}}, \bibinfo {author}
  {\bibfnamefont {M.~S.}\ \bibnamefont {Allman}}, \bibinfo {author}
  {\bibfnamefont {K.}~\bibnamefont {Cicak}}, \bibinfo {author} {\bibfnamefont
  {A.~J.}\ \bibnamefont {Sirois}}, \bibinfo {author} {\bibfnamefont {J.~D.}\
  \bibnamefont {Whittaker}}, \bibinfo {author} {\bibfnamefont {K.~W.}\
  \bibnamefont {Lehnert}}, \ and\ \bibinfo {author} {\bibfnamefont {R.~W.}\
  \bibnamefont {Simmonds}},\ }\href@noop {} {\bibfield  {journal} {\bibinfo
  {journal} {Nature}\ }\textbf {\bibinfo {volume} {475}},\ \bibinfo {pages}
  {359} (\bibinfo {year} {2011})}\BibitemShut {NoStop}%
\bibitem [{\citenamefont {Chan}\ \emph {et~al.}(2011)\citenamefont {Chan},
  \citenamefont {Mayer~Alegre}, \citenamefont {Safavi-Naeini}, \citenamefont
  {Hill}, \citenamefont {Krause}, \citenamefont {Groblacher}, \citenamefont
  {Aspelmeyer},\ and\ \citenamefont {Painter}}]{chan2011}%
  \BibitemOpen
  \bibfield  {author} {\bibinfo {author} {\bibfnamefont {J.}~\bibnamefont
  {Chan}}, \bibinfo {author} {\bibfnamefont {T.~P.}\ \bibnamefont
  {Mayer~Alegre}}, \bibinfo {author} {\bibfnamefont {A.~H.}\ \bibnamefont
  {Safavi-Naeini}}, \bibinfo {author} {\bibfnamefont {J.~T.}\ \bibnamefont
  {Hill}}, \bibinfo {author} {\bibfnamefont {A.}~\bibnamefont {Krause}},
  \bibinfo {author} {\bibfnamefont {S.}~\bibnamefont {Groblacher}}, \bibinfo
  {author} {\bibfnamefont {M.}~\bibnamefont {Aspelmeyer}}, \ and\ \bibinfo
  {author} {\bibfnamefont {O.}~\bibnamefont {Painter}},\ }\href@noop {}
  {\bibfield  {journal} {\bibinfo  {journal} {Nature}\ }\textbf {\bibinfo
  {volume} {478}},\ \bibinfo {pages} {89} (\bibinfo {year} {2011})}\BibitemShut
  {NoStop}%
\bibitem [{\citenamefont {Teufel}\ \emph {et~al.}(2009)\citenamefont {Teufel},
  \citenamefont {Donner}, \citenamefont {Castellanos-Beltran}, \citenamefont
  {Harlow},\ and\ \citenamefont {Lehnert}}]{teufel2009}%
  \BibitemOpen
  \bibfield  {author} {\bibinfo {author} {\bibfnamefont {J.~D.}\ \bibnamefont
  {Teufel}}, \bibinfo {author} {\bibfnamefont {T.}~\bibnamefont {Donner}},
  \bibinfo {author} {\bibfnamefont {M.~A.}\ \bibnamefont
  {Castellanos-Beltran}}, \bibinfo {author} {\bibfnamefont {J.~W.}\
  \bibnamefont {Harlow}}, \ and\ \bibinfo {author} {\bibfnamefont {K.~W.}\
  \bibnamefont {Lehnert}},\ }\href@noop {} {\bibfield  {journal} {\bibinfo
  {journal} {Nature Nanotech.}\ }\textbf {\bibinfo {volume} {4}},\ \bibinfo
  {pages} {820} (\bibinfo {year} {2009})}\BibitemShut {NoStop}%
\bibitem [{\citenamefont {Anetsberger}\ \emph {et~al.}(2010)\citenamefont
  {Anetsberger}, \citenamefont {Gavartin}, \citenamefont {Arcizet},
  \citenamefont {Unterreithmeier}, \citenamefont {Weig}, \citenamefont
  {Gorodetsky}, \citenamefont {Kotthaus},\ and\ \citenamefont
  {Kippenberg}}]{anetsberger2010}%
  \BibitemOpen
  \bibfield  {author} {\bibinfo {author} {\bibfnamefont {G.}~\bibnamefont
  {Anetsberger}}, \bibinfo {author} {\bibfnamefont {E.}~\bibnamefont
  {Gavartin}}, \bibinfo {author} {\bibfnamefont {O.}~\bibnamefont {Arcizet}},
  \bibinfo {author} {\bibfnamefont {Q.~P.}\ \bibnamefont {Unterreithmeier}},
  \bibinfo {author} {\bibfnamefont {E.~M.}\ \bibnamefont {Weig}}, \bibinfo
  {author} {\bibfnamefont {M.~L.}\ \bibnamefont {Gorodetsky}}, \bibinfo
  {author} {\bibfnamefont {J.~P.}\ \bibnamefont {Kotthaus}}, \ and\ \bibinfo
  {author} {\bibfnamefont {T.~J.}\ \bibnamefont {Kippenberg}},\ }\href@noop {}
  {\bibfield  {journal} {\bibinfo  {journal} {Phys. Rev. A}\ }\textbf {\bibinfo
  {volume} {82}},\ \bibinfo {pages} {061804(R)} (\bibinfo {year}
  {2010})}\BibitemShut {NoStop}%
\bibitem [{\citenamefont {Rugar}\ and\ \citenamefont
  {Gr\"{u}tter}(1991)}]{rugar1991}%
  \BibitemOpen
  \bibfield  {author} {\bibinfo {author} {\bibfnamefont {R.}~\bibnamefont
  {Rugar}}\ and\ \bibinfo {author} {\bibfnamefont {P.}~\bibnamefont
  {Gr\"{u}tter}},\ }\href@noop {} {\bibfield  {journal} {\bibinfo  {journal}
  {Phys. Rev. Lett.}\ }\textbf {\bibinfo {volume} {67}},\ \bibinfo {pages}
  {699} (\bibinfo {year} {1991})}\BibitemShut {NoStop}%
\bibitem [{\citenamefont {Natarajan}\ \emph {et~al.}(1995)\citenamefont
  {Natarajan}, \citenamefont {DiFilippo},\ and\ \citenamefont
  {Pritchard}}]{natarajan1995}%
  \BibitemOpen
  \bibfield  {author} {\bibinfo {author} {\bibfnamefont {V.}~\bibnamefont
  {Natarajan}}, \bibinfo {author} {\bibfnamefont {F.}~\bibnamefont
  {DiFilippo}}, \ and\ \bibinfo {author} {\bibfnamefont {D.~E.}\ \bibnamefont
  {Pritchard}},\ }\href@noop {} {\bibfield  {journal} {\bibinfo  {journal}
  {Phys. Rev. Lett.}\ }\textbf {\bibinfo {volume} {74}},\ \bibinfo {pages}
  {2855} (\bibinfo {year} {1995})}\BibitemShut {NoStop}%
\bibitem [{\citenamefont {Karabalin}\ \emph {et~al.}(2011)\citenamefont
  {Karabalin}, \citenamefont {Lifshitz}, \citenamefont {Cross}, \citenamefont
  {Matheny}, \citenamefont {Masmanidis},\ and\ \citenamefont
  {Roukes}}]{karabalin2011}%
  \BibitemOpen
  \bibfield  {author} {\bibinfo {author} {\bibfnamefont {R.~B.}\ \bibnamefont
  {Karabalin}}, \bibinfo {author} {\bibfnamefont {R.}~\bibnamefont {Lifshitz}},
  \bibinfo {author} {\bibfnamefont {M.~C.}\ \bibnamefont {Cross}}, \bibinfo
  {author} {\bibfnamefont {M.~H.}\ \bibnamefont {Matheny}}, \bibinfo {author}
  {\bibfnamefont {S.~C.}\ \bibnamefont {Masmanidis}}, \ and\ \bibinfo {author}
  {\bibfnamefont {M.~L.}\ \bibnamefont {Roukes}},\ }\href@noop {} {\bibfield
  {journal} {\bibinfo  {journal} {Phys. Rev. Lett.}\ }\textbf {\bibinfo
  {volume} {106}},\ \bibinfo {pages} {094102} (\bibinfo {year}
  {2011})}\BibitemShut {NoStop}%
\bibitem [{\citenamefont {Faust}\ \emph {et~al.}(2013)\citenamefont {Faust},
  \citenamefont {Rieger}, \citenamefont {Seitner}, \citenamefont {Kotthaus},\
  and\ \citenamefont {Weig}}]{faust2013}%
  \BibitemOpen
  \bibfield  {author} {\bibinfo {author} {\bibfnamefont {T.}~\bibnamefont
  {Faust}}, \bibinfo {author} {\bibfnamefont {J.}~\bibnamefont {Rieger}},
  \bibinfo {author} {\bibfnamefont {M.~J.}\ \bibnamefont {Seitner}}, \bibinfo
  {author} {\bibfnamefont {J.~P.}\ \bibnamefont {Kotthaus}}, \ and\ \bibinfo
  {author} {\bibfnamefont {E.~M.}\ \bibnamefont {Weig}},\ }\href@noop {}
  {\bibfield  {journal} {\bibinfo  {journal} {Nature Phys.}\ }\textbf {\bibinfo
  {volume} {9}},\ \bibinfo {pages} {485} (\bibinfo {year} {2013})}\BibitemShut
  {NoStop}%
\bibitem [{\citenamefont {Mahboob}\ \emph {et~al.}(2014)\citenamefont
  {Mahboob}, \citenamefont {Okamoto}, \citenamefont {Onomitsu},\ and\
  \citenamefont {Yamaguchi}}]{mahboob2014}%
  \BibitemOpen
  \bibfield  {author} {\bibinfo {author} {\bibfnamefont {I.}~\bibnamefont
  {Mahboob}}, \bibinfo {author} {\bibfnamefont {H.}~\bibnamefont {Okamoto}},
  \bibinfo {author} {\bibfnamefont {K.}~\bibnamefont {Onomitsu}}, \ and\
  \bibinfo {author} {\bibfnamefont {H.}~\bibnamefont {Yamaguchi}},\ }\href@noop
  {} {\bibfield  {journal} {\bibinfo  {journal} {arXiv:1405.5270}\ } (\bibinfo
  {year} {2014})}\BibitemShut {NoStop}%
\bibitem [{\citenamefont {Seok}\ \emph {et~al.}(2012)\citenamefont {Seok},
  \citenamefont {Buchmann}, \citenamefont {Singh},\ and\ \citenamefont
  {Meystre}}]{seok2012}%
  \BibitemOpen
  \bibfield  {author} {\bibinfo {author} {\bibfnamefont {H.}~\bibnamefont
  {Seok}}, \bibinfo {author} {\bibfnamefont {L.~F.}\ \bibnamefont {Buchmann}},
  \bibinfo {author} {\bibfnamefont {S.}~\bibnamefont {Singh}}, \ and\ \bibinfo
  {author} {\bibfnamefont {P.}~\bibnamefont {Meystre}},\ }\href@noop {}
  {\bibfield  {journal} {\bibinfo  {journal} {Phys. Rev. A}\ }\textbf {\bibinfo
  {volume} {86}},\ \bibinfo {pages} {063829} (\bibinfo {year}
  {2012})}\BibitemShut {NoStop}%
\bibitem [{\citenamefont {Tan}\ \emph {et~al.}(2013)\citenamefont {Tan},
  \citenamefont {Li},\ and\ \citenamefont {Meystre}}]{tan2013}%
  \BibitemOpen
  \bibfield  {author} {\bibinfo {author} {\bibfnamefont {H.}~\bibnamefont
  {Tan}}, \bibinfo {author} {\bibfnamefont {G.}~\bibnamefont {Li}}, \ and\
  \bibinfo {author} {\bibfnamefont {P.}~\bibnamefont {Meystre}},\ }\href@noop
  {} {\bibfield  {journal} {\bibinfo  {journal} {Phys. Rev. A}\ }\textbf
  {\bibinfo {volume} {87}},\ \bibinfo {pages} {033829} (\bibinfo {year}
  {2013})}\BibitemShut {NoStop}%
\bibitem [{\citenamefont {Woolley}\ and\ \citenamefont
  {Clerk}(2014)}]{woolley2014}%
  \BibitemOpen
  \bibfield  {author} {\bibinfo {author} {\bibfnamefont {M.}~\bibnamefont
  {Woolley}}\ and\ \bibinfo {author} {\bibfnamefont {A.~A.}\ \bibnamefont
  {Clerk}},\ }\href@noop {} {\bibfield  {journal} {\bibinfo  {journal} {Phys.
  Rev. A}\ }\textbf {\bibinfo {volume} {89}} (\bibinfo {year}
  {2014})}\BibitemShut {NoStop}%
\bibitem [{\citenamefont {O'Connell}\ \emph {et~al.}(2010)\citenamefont
  {O'Connell}, \citenamefont {Hofheinz}, \citenamefont {Ansmann}, \citenamefont
  {Bialczak}, \citenamefont {Lenander}, \citenamefont {Lucero}, \citenamefont
  {Neeley}, \citenamefont {Sank}, \citenamefont {Wang}, \citenamefont {Weides},
  \citenamefont {Wenner}, \citenamefont {Martinis},\ and\ \citenamefont
  {Cleland}}]{oconnell2010}%
  \BibitemOpen
  \bibfield  {author} {\bibinfo {author} {\bibfnamefont {A.~D.}\ \bibnamefont
  {O'Connell}}, \bibinfo {author} {\bibfnamefont {M.}~\bibnamefont {Hofheinz}},
  \bibinfo {author} {\bibfnamefont {M.}~\bibnamefont {Ansmann}}, \bibinfo
  {author} {\bibfnamefont {R.~C.}\ \bibnamefont {Bialczak}}, \bibinfo {author}
  {\bibfnamefont {M.}~\bibnamefont {Lenander}}, \bibinfo {author}
  {\bibfnamefont {E.}~\bibnamefont {Lucero}}, \bibinfo {author} {\bibfnamefont
  {M.}~\bibnamefont {Neeley}}, \bibinfo {author} {\bibfnamefont
  {D.}~\bibnamefont {Sank}}, \bibinfo {author} {\bibfnamefont {H.}~\bibnamefont
  {Wang}}, \bibinfo {author} {\bibfnamefont {M.}~\bibnamefont {Weides}},
  \bibinfo {author} {\bibfnamefont {J.}~\bibnamefont {Wenner}}, \bibinfo
  {author} {\bibfnamefont {J.~M.}\ \bibnamefont {Martinis}}, \ and\ \bibinfo
  {author} {\bibfnamefont {A.~N.}\ \bibnamefont {Cleland}},\ }\href@noop {}
  {\bibfield  {journal} {\bibinfo  {journal} {Nature}\ }\textbf {\bibinfo
  {volume} {464}},\ \bibinfo {pages} {697} (\bibinfo {year}
  {2010})}\BibitemShut {NoStop}%
\bibitem [{\citenamefont {Suh}\ \emph {et~al.}(2010)\citenamefont {Suh},
  \citenamefont {LaHaye}, \citenamefont {Echternach}, \citenamefont {Schwab},\
  and\ \citenamefont {Roukes}}]{suh2010}%
  \BibitemOpen
  \bibfield  {author} {\bibinfo {author} {\bibfnamefont {J.}~\bibnamefont
  {Suh}}, \bibinfo {author} {\bibfnamefont {M.~D.}\ \bibnamefont {LaHaye}},
  \bibinfo {author} {\bibfnamefont {P.~M.}\ \bibnamefont {Echternach}},
  \bibinfo {author} {\bibfnamefont {K.~C.}\ \bibnamefont {Schwab}}, \ and\
  \bibinfo {author} {\bibfnamefont {M.~L.}\ \bibnamefont {Roukes}},\
  }\href@noop {} {\bibfield  {journal} {\bibinfo  {journal} {Nano Lett.}\
  }\textbf {\bibinfo {volume} {10}},\ \bibinfo {pages} {3990} (\bibinfo {year}
  {2010})}\BibitemShut {NoStop}%
\bibitem [{\citenamefont {Hertzberg}\ \emph {et~al.}(2010)\citenamefont
  {Hertzberg}, \citenamefont {Rocheleau}, \citenamefont {Ndukum}, \citenamefont
  {Savva}, \citenamefont {Clerk},\ and\ \citenamefont {Schwab}}]{hertz2010}%
  \BibitemOpen
  \bibfield  {author} {\bibinfo {author} {\bibfnamefont {J.~B.}\ \bibnamefont
  {Hertzberg}}, \bibinfo {author} {\bibfnamefont {T.}~\bibnamefont
  {Rocheleau}}, \bibinfo {author} {\bibfnamefont {T.}~\bibnamefont {Ndukum}},
  \bibinfo {author} {\bibfnamefont {M.}~\bibnamefont {Savva}}, \bibinfo
  {author} {\bibfnamefont {A.~A.}\ \bibnamefont {Clerk}}, \ and\ \bibinfo
  {author} {\bibfnamefont {K.~C.}\ \bibnamefont {Schwab}},\ }\href@noop {}
  {\bibfield  {journal} {\bibinfo  {journal} {Nature Phys.}\ }\textbf {\bibinfo
  {volume} {6}},\ \bibinfo {pages} {213} (\bibinfo {year} {2010})}\BibitemShut
  {NoStop}%
\bibitem [{\citenamefont {Suh}\ \emph {et~al.}(2014)\citenamefont {Suh},
  \citenamefont {Weinstein}, \citenamefont {Lei}, \citenamefont {Wollman},
  \citenamefont {Steinke}, \citenamefont {Meystre}, \citenamefont {Clerk},\
  and\ \citenamefont {Schwab}}]{suh2014}%
  \BibitemOpen
  \bibfield  {author} {\bibinfo {author} {\bibfnamefont {J.}~\bibnamefont
  {Suh}}, \bibinfo {author} {\bibfnamefont {A.~J.}\ \bibnamefont {Weinstein}},
  \bibinfo {author} {\bibfnamefont {C.~U.}\ \bibnamefont {Lei}}, \bibinfo
  {author} {\bibfnamefont {E.~E.}\ \bibnamefont {Wollman}}, \bibinfo {author}
  {\bibfnamefont {S.~K.}\ \bibnamefont {Steinke}}, \bibinfo {author}
  {\bibfnamefont {P.}~\bibnamefont {Meystre}}, \bibinfo {author} {\bibfnamefont
  {A.~A.}\ \bibnamefont {Clerk}}, \ and\ \bibinfo {author} {\bibfnamefont
  {K.~C.}\ \bibnamefont {Schwab}},\ }\href@noop {} {\bibfield  {journal}
  {\bibinfo  {journal} {Science}\ }\textbf {\bibinfo {volume} {344}},\ \bibinfo
  {pages} {1262} (\bibinfo {year} {2014})}\BibitemShut {NoStop}%
\bibitem [{\citenamefont {Szorkovszky}\ \emph {et~al.}(2013)\citenamefont
  {Szorkovszky}, \citenamefont {Brawley}, \citenamefont {Doherty},\ and\
  \citenamefont {Bowen}}]{szork2013}%
  \BibitemOpen
  \bibfield  {author} {\bibinfo {author} {\bibfnamefont {A.}~\bibnamefont
  {Szorkovszky}}, \bibinfo {author} {\bibfnamefont {G.~A.}\ \bibnamefont
  {Brawley}}, \bibinfo {author} {\bibfnamefont {A.~C.}\ \bibnamefont
  {Doherty}}, \ and\ \bibinfo {author} {\bibfnamefont {W.~P.}\ \bibnamefont
  {Bowen}},\ }\href@noop {} {\bibfield  {journal} {\bibinfo  {journal} {Phys.
  Rev. Lett.}\ }\textbf {\bibinfo {volume} {110}},\ \bibinfo {pages} {184301}
  (\bibinfo {year} {2013})}\BibitemShut {NoStop}%
\bibitem [{\citenamefont {Vinante}\ and\ \citenamefont
  {Falferi}(2013)}]{vinante2013}%
  \BibitemOpen
  \bibfield  {author} {\bibinfo {author} {\bibfnamefont {A.}~\bibnamefont
  {Vinante}}\ and\ \bibinfo {author} {\bibfnamefont {P.}~\bibnamefont
  {Falferi}},\ }\href@noop {} {\bibfield  {journal} {\bibinfo  {journal} {Phys.
  Rev. Lett.}\ }\textbf {\bibinfo {volume} {111}},\ \bibinfo {pages} {207203}
  (\bibinfo {year} {2013})}\BibitemShut {NoStop}%
\bibitem [{\citenamefont {Szorkovszky}\ \emph {et~al.}(2014)\citenamefont
  {Szorkovszky}, \citenamefont {Clerk}, \citenamefont {Doherty},\ and\
  \citenamefont {Bowen}}]{szork2014}%
  \BibitemOpen
  \bibfield  {author} {\bibinfo {author} {\bibfnamefont {A.}~\bibnamefont
  {Szorkovszky}}, \bibinfo {author} {\bibfnamefont {A.~A.}\ \bibnamefont
  {Clerk}}, \bibinfo {author} {\bibfnamefont {A.~C.}\ \bibnamefont {Doherty}},
  \ and\ \bibinfo {author} {\bibfnamefont {W.~P.}\ \bibnamefont {Bowen}},\
  }\href@noop {} {\bibfield  {journal} {\bibinfo  {journal} {New J. Phys.}\
  }\textbf {\bibinfo {volume} {16}},\ \bibinfo {pages} {043023} (\bibinfo
  {year} {2014})}\BibitemShut {NoStop}%
\bibitem [{\citenamefont {Pontin}\ \emph {et~al.}(2014)\citenamefont {Pontin},
  \citenamefont {Bonaldi}, \citenamefont {Borrielli}, \citenamefont
  {Cataliotti}, \citenamefont {Marino}, \citenamefont {Prodi}, \citenamefont
  {Serra},\ and\ \citenamefont {Marin}}]{pontin2014}%
  \BibitemOpen
  \bibfield  {author} {\bibinfo {author} {\bibfnamefont {A.}~\bibnamefont
  {Pontin}}, \bibinfo {author} {\bibfnamefont {M.}~\bibnamefont {Bonaldi}},
  \bibinfo {author} {\bibfnamefont {A.}~\bibnamefont {Borrielli}}, \bibinfo
  {author} {\bibfnamefont {F.~S.}\ \bibnamefont {Cataliotti}}, \bibinfo
  {author} {\bibfnamefont {F.}~\bibnamefont {Marino}}, \bibinfo {author}
  {\bibfnamefont {G.~A.}\ \bibnamefont {Prodi}}, \bibinfo {author}
  {\bibfnamefont {E.}~\bibnamefont {Serra}}, \ and\ \bibinfo {author}
  {\bibfnamefont {F.}~\bibnamefont {Marin}},\ }\href@noop {} {\bibfield
  {journal} {\bibinfo  {journal} {Phys. Rev. A}\ }\textbf {\bibinfo {volume}
  {89}},\ \bibinfo {pages} {023848} (\bibinfo {year} {2014})}\BibitemShut
  {NoStop}%
\bibitem [{\citenamefont {Chakram}\ \emph {et~al.}(2014)\citenamefont
  {Chakram}, \citenamefont {Patil}, \citenamefont {Chang},\ and\ \citenamefont
  {Vengalattore}}]{chakram2014}%
  \BibitemOpen
  \bibfield  {author} {\bibinfo {author} {\bibfnamefont {S.}~\bibnamefont
  {Chakram}}, \bibinfo {author} {\bibfnamefont {Y.~S.}\ \bibnamefont {Patil}},
  \bibinfo {author} {\bibfnamefont {L.}~\bibnamefont {Chang}}, \ and\ \bibinfo
  {author} {\bibfnamefont {M.}~\bibnamefont {Vengalattore}},\ }\href@noop {}
  {\bibfield  {journal} {\bibinfo  {journal} {Phys. Rev. Lett.}\ }\textbf
  {\bibinfo {volume} {112}},\ \bibinfo {pages} {127201} (\bibinfo {year}
  {2014})}\BibitemShut {NoStop}%
\bibitem [{\citenamefont {Purdy}\ \emph {et~al.}(2014)\citenamefont {Purdy},
  \citenamefont {Yu}, \citenamefont {Kampel}, \citenamefont {Peterson},
  \citenamefont {Cicak}, \citenamefont {Simmonds},\ and\ \citenamefont
  {Regal}}]{purdy2014}%
  \BibitemOpen
  \bibfield  {author} {\bibinfo {author} {\bibfnamefont {T.~P.}\ \bibnamefont
  {Purdy}}, \bibinfo {author} {\bibfnamefont {P.~L.}\ \bibnamefont {Yu}},
  \bibinfo {author} {\bibfnamefont {N.~S.}\ \bibnamefont {Kampel}}, \bibinfo
  {author} {\bibfnamefont {R.~W.}\ \bibnamefont {Peterson}}, \bibinfo {author}
  {\bibfnamefont {K.}~\bibnamefont {Cicak}}, \bibinfo {author} {\bibfnamefont
  {R.~W.}\ \bibnamefont {Simmonds}}, \ and\ \bibinfo {author} {\bibfnamefont
  {C.~A.}\ \bibnamefont {Regal}},\ }\href@noop {} {\bibfield  {journal}
  {\bibinfo  {journal} {arXiv:1406.7241}\ } (\bibinfo {year}
  {2014})}\BibitemShut {NoStop}%
\bibitem [{\citenamefont {Lee}\ \emph {et~al.}(2014)\citenamefont {Lee},
  \citenamefont {Underwood}, \citenamefont {Mason}, \citenamefont {Shkarin},
  \citenamefont {Borkje}, \citenamefont {Girvin},\ and\ \citenamefont
  {Harris}}]{lee2014}%
  \BibitemOpen
  \bibfield  {author} {\bibinfo {author} {\bibfnamefont {D.}~\bibnamefont
  {Lee}}, \bibinfo {author} {\bibfnamefont {M.}~\bibnamefont {Underwood}},
  \bibinfo {author} {\bibfnamefont {D.}~\bibnamefont {Mason}}, \bibinfo
  {author} {\bibfnamefont {A.~B.}\ \bibnamefont {Shkarin}}, \bibinfo {author}
  {\bibfnamefont {K.}~\bibnamefont {Borkje}}, \bibinfo {author} {\bibfnamefont
  {S.~M.}\ \bibnamefont {Girvin}}, \ and\ \bibinfo {author} {\bibfnamefont
  {J.~G.~E.}\ \bibnamefont {Harris}},\ }\href@noop {} {\bibfield  {journal}
  {\bibinfo  {journal} {arXiv:1406.7254}\ } (\bibinfo {year}
  {2014})}\BibitemShut {NoStop}%
\bibitem [{\citenamefont {Gavartin}\ \emph {et~al.}(2013)\citenamefont
  {Gavartin}, \citenamefont {Verlot},\ and\ \citenamefont
  {Kippenberg}}]{gavartin2013}%
  \BibitemOpen
  \bibfield  {author} {\bibinfo {author} {\bibfnamefont {E.}~\bibnamefont
  {Gavartin}}, \bibinfo {author} {\bibfnamefont {P.}~\bibnamefont {Verlot}}, \
  and\ \bibinfo {author} {\bibfnamefont {T.~J.}\ \bibnamefont {Kippenberg}},\
  }\href@noop {} {\bibfield  {journal} {\bibinfo  {journal} {Nat. Commun.}\
  }\textbf {\bibinfo {volume} {4:2860}} (\bibinfo {year} {2013})}\BibitemShut
  {NoStop}%
\bibitem [{\citenamefont {Reid}\ and\ \citenamefont
  {Drummond}(1988)}]{reid1988}%
  \BibitemOpen
  \bibfield  {author} {\bibinfo {author} {\bibfnamefont {M.~D.}\ \bibnamefont
  {Reid}}\ and\ \bibinfo {author} {\bibfnamefont {P.~D.}\ \bibnamefont
  {Drummond}},\ }\href@noop {} {\bibfield  {journal} {\bibinfo  {journal}
  {Phys. Rev. Lett.}\ }\textbf {\bibinfo {volume} {60}},\ \bibinfo {pages}
  {2731} (\bibinfo {year} {1988})}\BibitemShut {NoStop}%
\bibitem [{\citenamefont {Lifshitz}\ and\ \citenamefont
  {Cross}(2010)}]{lifshitz2010}%
  \BibitemOpen
  \bibfield  {author} {\bibinfo {author} {\bibfnamefont {R.}~\bibnamefont
  {Lifshitz}}\ and\ \bibinfo {author} {\bibfnamefont {M.~C.}\ \bibnamefont
  {Cross}},\ }in\ \href@noop {} {\emph {\bibinfo {booktitle} {Reviews of
  nonlinear dynamics and complexity}}},\ \bibinfo {editor} {edited by\ \bibinfo
  {editor} {\bibfnamefont {H.~G.}\ \bibnamefont {Schuster}}}\ (\bibinfo {year}
  {2010})\ p.~\bibinfo {pages} {1}\BibitemShut {NoStop}%
\bibitem [{\citenamefont {Chakram}\ \emph {et~al.}()\citenamefont {Chakram},
  \citenamefont {Patil}, \citenamefont {Chang},\ and\ \citenamefont
  {Vengalattore}}]{imaging2013}%
  \BibitemOpen
  \bibfield  {author} {\bibinfo {author} {\bibfnamefont {S.}~\bibnamefont
  {Chakram}}, \bibinfo {author} {\bibfnamefont {Y.~S.}\ \bibnamefont {Patil}},
  \bibinfo {author} {\bibfnamefont {L.}~\bibnamefont {Chang}}, \ and\ \bibinfo
  {author} {\bibfnamefont {M.}~\bibnamefont {Vengalattore}},\ }\href@noop {}
  {\bibinfo  {journal} {(to be published)}\ }\BibitemShut {NoStop}%
\end{thebibliography}%

\section*{Supplementary Information}
\subsection*{Photothermal frequency stabilization}
The mechanical modes of the membrane are susceptible to large drifts in frequency due to temperature fluctuations and the differential expansion between the silicon substrate and the Silicon nitride membrane. We measure drifts on the order of 500 Hz ($10^3$ - $10^4 \, \gamma$) per Kelvin for the modes studied in this work. Thus, resolving thermomechanical motion and non-thermal two-mode correlations requires active sub-linewidth stabilization of the mechanical eigenfrequencies.  

The frequencies of the mechanical modes are stabilized by photothermal control of the silicon substrate. Based on our observation that temperature fluctuations cause frequency drifts that are highly correlated across the various modes of the membrane, we implement active stabilization by continuously monitoring the frequency of a high-$Q$ membrane mode at 2.736 MHz - far from the modes that exhibit the two-mode nonlinearity studied in this work. This `thermometer mode' has a quality factor in excess of $8 \times 10^6$ ($\gamma/2 \pi < 340$ mHz). Phase sensitive detection of this mode results in an error signal with an on-resonant phase slope of 5.91 radians/Hz. Active photothermal stabilization of the substrate is accomplished with typical optical powers of 600 $\mu$W, leading to rms frequency fluctuations of this thermometer mode below 2 mHz (equivalent to temperature fluctuations of the substrate of less than 2 $\mu$K). For the mechanical modes relevant to this work, this stabilization translates to frequency fluctuations less than $0.002 \times \gamma$. 

\subsection*{Nondegenerate parametric amplifier below threshold : Phase-sensitive amplification}
We model the nonlinear interaction between membrane modes by a Hamiltonian given by $\mathcal{H}_{ij} = - g X_S x_i x_j$ where $g$ is the interaction strength, $X_S$ is the displacement of the substrate and $x_{i,j}$ represent the displacement of the membrane modes. Within the rotating wave approximation, this results in equations of motion of the form
\begin{equation}
\ddot{z}_i + \gamma_i \dot{z}_i + \omega_i^2 z_i = \frac{1}{m_i} (F_i(t) + \frac{g}{2} Z_S z^*_j)
\end{equation}
and similar equations for the other membrane mode and the substrate. Here, $z_i$ is the (complex) displacement of the mode, $F_i$ represents both the classical actuation as well as thermomechanical noise forces and $\omega, \gamma$ are the eigenfrequencies and mechanical linewidths. The coupled equations of motion can be solved using the methods of two time scale perturbation theory \cite{lifshitz2010}. This gives rise to coupled equations of the form 
\begin{eqnarray}
2 \dot{A}_{i}  &=& \gamma_{i} \left[ - A_{i} + i \chi_{i} \left( \frac{g}{2} A^*_{j} A_S + \tilde{F_{i}} \right) \right] \\
2 \dot{A}_{j}  &=& \gamma_{j} \left[ - A_{j} + i \chi_{j} \left( \frac{g}{2} A^*_{i} A_S + \tilde{F_{j}} \right) \right]
\end{eqnarray}
where $z_i = A_i e^{-i \omega_i t}$ and we have ignored terms such as $\ddot{A}_i, \gamma_i \dot{A}_i$ in the slow time approximation and $\tilde{F}_{i,j}$ are the slowly varying (complex) amplitudes of the external forces on the individual membrane modes. Even in the absence of external forces $\tilde{F_{i,j}}$, these coupled equations allow for non-zero steady state amplitudes, i.e. parametric self-oscillation, above a threshold substrate amplitude given by
\begin{equation}
X_{S, th} = 2\ \sqrt{\frac{1}{g^2}{ \frac{1}{\chi_i \chi_j}}}
\end{equation}
While we focus on the below-threshold behavior of this nondegenerate parametric amplifier in this work, these coupled equations also accurately describe various nonlinear aspects of the amplifier above threshold \cite{imaging2013}. 

In the presence of external actuation of the individual membrane modes and the substrate (`pump') mode, i.e. $\tilde{F_{i,j}} \neq 0, A_S \neq 0$, the amplitude of each mode is a coherent superposition of its individual response to the external force and the down-converted phonons arising from the two-mode nonlinearity. Thus, the complex amplitudes in steady state are given by
\begin{eqnarray}
A_{i} &=& \frac{e^{i(\phi_i - \pi/2)}}{1 - \mu^2} \left( \chi_i |\tilde{F}_i| + \mu \sqrt{\chi_i \chi_j} |\tilde{F}_{j}| e^{i \delta \phi} \right) \\
A_{j} &=& \frac{e^{i(\phi_j - \pi/2)}}{1 - \mu^2} \left( \chi_j |\tilde{F}_j| + \mu \sqrt{\chi_i \chi_j} |\tilde{F}_{i}| e^{i \delta \phi} \right)
\end{eqnarray}
where $\mu = X_S/X_{S,th}$ is the pump amplitude normalized to the threshold for parametric instability and $\delta \phi = \phi_S - \phi_i - \phi_j$ with $\phi_{i,j,S}$ being the various phases associated with the external forces. 

The above equation can be recast in terms of a phase-dependent gain $G_{i,j}(\delta \phi) = |A_{i,j}|/|A^0_{i,j}|$ where $A^0_{i,j}$ are the steady state amplitudes of the respective modes in the absence of down-conversion. This phase-dependent gain is then given by
\begin{eqnarray}
G_{j}(\delta \phi) &=& \frac{1}{1 - \mu^2}  \nonumber \\
	&\times& \sqrt{1 + \mu^2 \frac{\chi_i}{\chi_j} \left(\frac{|\tilde{F}_{i}|}{|\tilde{F}_{j}|}\right)^2 - 2 \mu \left(\frac{\chi_i}{\chi_j}\right)^{1/2} \frac{|\tilde{F}_{i}|}{|\tilde{F}_{j}|}\cos(\delta \phi)} \nonumber \\
	&=& \frac{1}{1 - \mu^2} \sqrt{1 + \mu^2 \eta^2 - 2 \mu \eta \cos(\delta \phi)}
\end{eqnarray}
where $\eta = (\chi_j/\chi_i)^{1/2} \times (\bar{x}_i/\bar{x}_j)$ and $\bar{x}_{i,j}$ are the amplitudes of the membrane modes in the absence of the pump. A no-free-parameter fit of this expression is in agreement with our data to within 5\% (Fig. 3).

\begin{suppfig}[t]
\includegraphics[width=2.70in]{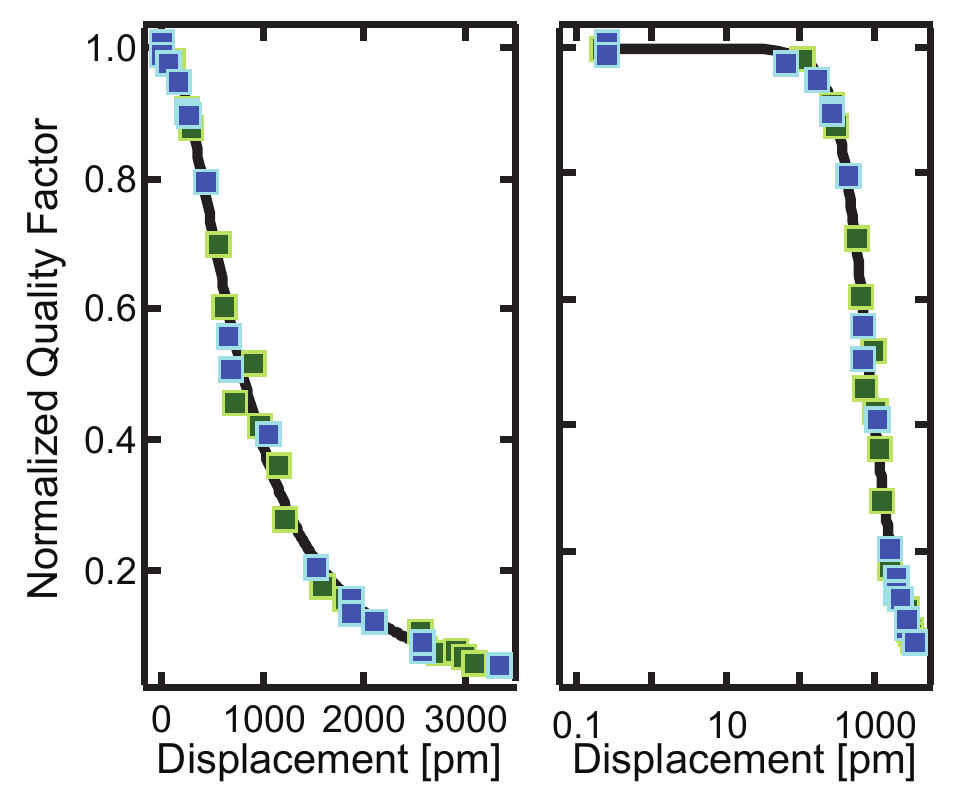}
\caption{Two-mode dissipation due to upconversion of energy into the substrate. The normalized quality factors $Q(x)/Q_0$ are shown for the measured dissipation of mode $i$ due to actuation of mode $j$ (green) and vice versa (blue), as a function of displacement in linear (left) and log (right) scale. For comparison, typical amplitudes of thermomechanical motion are in the range 0.1-0.2 pm. Solid lines show the fits of the data to the two-mode model.}
\label{Fig:FigS1}
\end{suppfig}
 
\subsection*{Two-mode control of mechanical dissipation}

For sufficiently large amplitudes of membrane motion, the two-mode nonlinearity can result in the coherent upconversion of excitations into the substrate. In our system, we typically measure quality factors of $Q_S \sim 10^3 - 10^4$ for the substrate modes (e.g. flexural modes), about three orders of magnitude lower than those of the membrane. 
Due to the relatively large dissipation rate of substrate modes ($\gamma_S/\gamma_{i,j} > 10^3$), this upconversion process can be regarded as nonlinear two-mode dissipation where the dissipation rate of a mode $i$ is influenced by the motion of its partner mode $j$. 

The modified dissipation rate $\gamma_i(x_j)$ of a mode $i$ can be most easily seen by writing the coupled equations of motion for the slowly varying amplitudes $A_{i, j, S}$. Here, we assume that mode $j$ is actuated externally to a large amplitude in the absence of substrate (pump) actuation. The coupled equations become
\begin{eqnarray}
2 \dot{A}_i + \gamma_i A_i &=& \gamma_i \chi_i \frac{g}{2} |A_j| A_S \\
2 \dot{A}_S + \gamma_S A_S &=& - \gamma_S \chi_S \frac{g}{2} |A_j| A_i 
\end{eqnarray}
where we assume that $A_j$ is purely imaginary. Diagonalizing these equations leads to a modified damping rate of mode $i$ given by
\begin{equation}
\gamma_i (\bar{x}_j) = \frac{1}{2} \left[ \gamma_S + \gamma_i - \sqrt{(\gamma_S - \gamma_i)^2 -  \gamma_S^2 \frac{\bar{x}_j^2}{\xi^2}  } \right]
\end{equation}
where $\xi = \frac{1}{2} (\gamma_S/\gamma_i)^{1/2} (\chi_j / \chi_S)^{1/2} \times X_{S, th}$. This can be recast in terms of a normalized quality factor
$Q_i(\bar{x}_j)/Q_{i,0} = (\gamma_i(\bar{x}_j)/\gamma_i)^{-1}$ where $Q_{i,0}$ is the quality factor of membrane mode $i$ in the absence of the two-mode nonlinearity.

Our data are in excellent agreement with this expression for a wide range of mode pairs that were seen to exhibit this two-mode nonlinearity (see, for example, Fig. S1). 

As can be seen in Fig. S1, a significant modification to the dissipation rate of a membrane mode $i$ requires large amplitude actuation of its partner mode $j$, typically over 4-5 orders of magnitude larger than thermomechanical amplitudes and much larger than the typical scale of motion for the studies presented in this work. Thus, while these measurements constitute an important validation of our two-mode model, the two-mode contribution to the individual dissipation rates are negligible for our studies on parametric amplification and thermomechanical squeezing. 

\subsection*{Thermomechanical two-mode squeezing}
The nondegenerate parametric amplifier, when driven below threshold, develops correlations between the non-degenerate modes. These correlations are manifest as a squeezing of a composite quadrature formed from linear combinations of the individual membrane modes. We analyze the coupled equations for the membrane mode under the influence of a classical actuation of the substrate mode below threshold. Furthermore, we assume that the membrane modes are subject to thermomechanical noise forces. The influence of such noise forces on the substrate mode can be neglected due to the large classical drive imposed on this mode. Finally, for each membrane mode, we distinguish between its mean amplitude and its fluctuations by writing $z_{i,j} = (\bar{A}_{i,j} + \delta A_{i,j}) e^{-i \omega_{i,j}t}$ where $\langle \delta A_{i,j} \rangle = 0$. The coupled equations for the fluctuations can be written as
%\begin{eqnarray}
\begin{align*}
&2 \left( \begin{array}{c}
	\delta \dot{A}_i \\ \delta \dot{A}_j \\ \delta \dot{A}_S \end{array} \right) &\\
&= \left( \begin{array}{ccc} -\gamma_i & 0 & i \gamma_i \chi_i \frac{g}{2} \bar{A}_j^* \\
	0 & -\gamma_j & i \gamma_j \chi_j \frac{g}{2} \bar{A}_i^* \\ i \gamma_S \chi_S \frac{g}{2} \bar{A}_j & i \gamma_S \chi_S \frac{g}{2} \bar{A}_i & -\gamma_S \end{array}\right) \left( \begin{array}{c}
	\delta A_i \\ \delta A_j \\ \delta A_S \end{array} \right) \nonumber &\\
&+  \left( \begin{array}{ccc} 0 & i \gamma_i \chi_i \frac{g}{2} \bar{A}_S & 0 \\
	i \gamma_j \chi_j \frac{g}{2} \bar{A}_S & 0 & 0 \\ 0 & 0 & 0 \end{array}\right) \left( \begin{array}{c}
	\delta A^*_i \\ \delta A^*_j \\ \delta A^*_S \end{array} \right) \nonumber &\\
&+ i \left( \begin{array}{c}
	\gamma_i \chi_i F_i \\ \gamma_j \chi_j F_j \\ \gamma_S \chi_S F_S \end{array} \right)&
\end{align*}
%\end{eqnarray}
where the thermomechanical noise forces obey
\begin{eqnarray}
\langle F_i(t) \rangle = \langle F_i (t) F_j (t') \rangle = 0, \\
\langle F_i (t) F^*_j (t+\tau) \rangle = 8 \gamma_i m_i k_B T \delta_{ij} \delta(\tau)
\end{eqnarray}
Using the decomposition
\begin{eqnarray}
\delta {\bf A} &=& \delta {\vec \alpha} + i \delta {\vec \beta} \\
\mathbf{v} &=& \mathbf{v}_{\alpha} + i\mathbf{v}_{\beta}
\end{eqnarray}
where $\delta {\bf A} = (\delta A_i, \delta A_j, \delta A_S)^T, \\ \mathbf{v} = \frac{i}{2}  (\gamma_{i}\chi_{i}F_{i},\gamma_{j}\chi_{j}F_{j},\gamma_{S}\chi_{S}F_{S})^{T}$, etc. leads to the following equations of motion
\begin{eqnarray}
\delta \dot{\vec \alpha} &=& {\bf M}_\alpha \delta {\vec \alpha} + {\bf v}_\alpha \\
\delta \dot{\vec \beta} &=& {\bf M}_\beta \delta {\vec \beta} + {\bf v}_\beta
\end{eqnarray}
where 
\begin{equation}
{\bf M}_{\alpha,\beta} = \frac{1}{2} \left( \begin{array}{ccc} -\gamma_i & \pm \gamma_i \chi_i \frac{g}{2} |\bar{A}_S| & \gamma_i \chi_i \frac{g}{2} |\bar{A}_j| \\
\pm \gamma_j \chi_j \frac{g}{2} |\bar{A}_S| & -\gamma_j &  - \gamma_j \chi_j \frac{g}{2} |\bar{A}_i| \\
- \gamma_S \chi_S \frac{g}{2} |\bar{A}_j| & \gamma_S \chi_S \frac{g}{2} |\bar{A}_i| & -\gamma_S \end{array}\right)
\end{equation}
and the elements of ${\bf v}_{\alpha,\beta}$ satisfy $\langle v_i \rangle = 0$, $\langle v_k (t) v_l (t + \tau) \rangle = \frac{\gamma_l k_B T}{m_l \omega_l^2}\delta_{kl} \delta (\tau)$. 
The spectrum in steady state is thus given by the matrix equation
\begin{equation}
{\bf S}_{\alpha, \beta}(\omega) = \frac{1}{2 \pi} ({\bf M}_{\alpha,\beta} + i \omega {\bf I})^{-1} {\bf D} ({\bf M}^T_{\alpha,\beta} - i \omega {\bf I})^{-1}
\end{equation}
where ${\bf I}$ is the identity and 
\begin{equation}
{\bf D}  = k_B T \left( \begin{array}{ccc} \frac{\gamma_i}{m_i \omega_i^2} & 0 & 0 \\ 0 & \frac{\gamma_j}{m_j \omega_j^2} & 0 \\ 0 & 0 & \frac{\gamma_S}{M \omega_S^2} \end{array}\right)
\end{equation}
is the diffusion matrix. 

The zero-time correlations in steady state can be obtained from the spectrum using the Wiener-Khintchine theorem, and are given by 
\begin{eqnarray}
\langle \delta \alpha_i \delta \alpha_i \rangle &=& \frac{k_B T}{m_i \omega_i^2} \frac{1}{1 - \mu^2} \left[ 1 - \delta \mu^2 + \mu^2 \frac{\gamma_j}{2 \bar{\gamma}} \left( \frac{\omega_i }{\omega_j} - 1 \right)\right] \nonumber \\
\langle \delta \alpha_j \delta \alpha_j \rangle &=& \frac{k_B T}{m_j \omega_j^2} \frac{1}{1 - \mu^2} \left[ 1 + \delta \mu^2 + \mu^2 \frac{\gamma_i}{2 \bar{\gamma}} \left( \frac{\omega_j }{\omega_i} - 1 \right)\right] \nonumber \\
\langle \delta \alpha_i \delta \alpha_j \rangle &=& \frac{k_B T}{\sqrt{m_i \omega_i^2}\sqrt{m_j \omega_j^2}} \left[ \frac{\mu (1 - \delta^2)^{1/2}}{2 (1 - \mu^2)} \left(\sqrt{\frac{\omega_i}{\omega_j}} + \sqrt{\frac{\omega_j}{\omega_i}} \right)\right] \nonumber
\end{eqnarray}
where $\delta = (\gamma_i - \gamma_j)/(\gamma_i + \gamma_j)$ is the `loss asymmetry' and $\bar{\gamma} = (\gamma_i + \gamma_j)/2$. 

In our measurements, we quantify the degree of two-mode squeezing by defining cross-quadratures constructed from $\{ \alpha_{i,j}, \beta_{i,j} \}$ normalized to their respective thermomechanical amplitudes, according to the relations $x_{a,b} = (\alpha_i \pm \alpha_j)/\sqrt{2}, y_{a,b} = (\beta_i \pm \beta_j)/\sqrt{2}$. The standard deviations of these cross-quadratures can be obtained from the above expressions. These are shown in Fig. 4 for the independently measured parameters of our nondegenerate parametric amplifier. These data are obtained with a measurement filter bandwidth of $\Delta \nu = 10$ Hz, i.e. $\Delta \nu > 10^2 \gamma_{i,j}$. Thus, the correlations calculated above by integrating over all frequencies (i.e. infinite bandwidth), are a good approximation to our measurements.  As can be expected, the degree of squeezing increases monotonically with decreasing filter bandwidth, reaching the 6 dB limit for $\Delta \nu < \gamma_{i,j}/2 \pi$ in the case of symmetric dissipation rates, i.e. $\delta = 0$ \, \cite{imaging2013}.

\end{document}